\documentclass[journal]{IEEEtran}

\usepackage{booktabs}
\usepackage{threeparttable}
\usepackage{amsmath,graphicx,cite,amssymb}
\usepackage{amsfonts,multirow,bm,array,setspace,stfloats}
\usepackage{xcolor}
\usepackage{xpatch}

\usepackage{graphicx,float,cite,amssymb}
\usepackage{graphics}
 \DeclareGraphicsExtensions{.pdf,.jpeg,.png,.jpg}   
\usepackage{multirow,bm,bbm,array,setspace}
\usepackage{textcomp}
\usepackage{psfrag}
\usepackage{color}
\usepackage{pstricks,enumerate}
\usepackage{bbm}
\usepackage{subfigure}

\usepackage{algorithm,algpseudocode}
\makeatletter

\def\changeBibColor#1{%
  \ifin@\color{blue}\else\normalcolor\fi
}
 
\xpatchcmd\@bibitem
  {\item}
  {\changeBibColor{#1}\item}
  {}{\fail}
 
\xpatchcmd\@lbibitem
  {\item}
  {\changeBibColor{#2}\item}
  {}{\fail}
\makeatother

\newcommand{\distas}[1]{\mathbin{\overset{#1}{\kern\z@\sim}}}%
\newsavebox{\mybox}\newsavebox{\mysim}
\newcommand{\distras}[1]{%
  \savebox{\mybox}{\hbox{\kern1pt$\scriptstyle#1$\kern1pt}}%
  \savebox{\mysim}{\hbox{$\sim$}}%
  \mathbin{\overset{#1}{\kern\z@\resizebox{\wd\mybox}{\ht\mysim}{$\sim$}}}%
}
\makeatother
\ifCLASSINFOpdf
\else
\fi

\newtheorem{proposition}{Proposition}

\newtheorem{lemma}{Lemma}
\newtheorem{theorem}{Theorem}

\newtheorem{remark}{Remark}

\begin{document}
%
\title{A Linear Time Algorithm for the Optimal\\ Discrete IRS Beamforming}
\author{
	\IEEEauthorblockN{
	Shuyi Ren, \IEEEmembership{Student Member,~IEEE}, 
	Kaiming Shen, \IEEEmembership{Member,~IEEE},\\
	Xin Li, Xin Chen, and Zhi-Quan Luo, \IEEEmembership{Fellow,~IEEE}
} 
\thanks{Accepted to IEEE Wireless Communications Letters on December 23, 2022. This work was supported in part by the National Natural Science Foundation of China (NSFC) under Grant 92167202, in part by the NSFC under Grant 62001411, in part by the Huawei Technologies, and in part by the Guangdong
Provincial Key Laboratory of Big Data Computing. \emph{(Corresponding author: Kaiming Shen.)}

S. Ren and K. Shen are with The Chinese University of Hong Kong, Shenzhen, 518172, China (e-mail: shuyiren@link.cuhk.edu.cn; shenkaiming@cuhk.edu.cn). 

X. Li and X. Chen are with Huawei Technologies (e-mail: razor.lixin@huawei.com).

Z.-Q. Luo is both with The Chinese University of Hong Kong, Shenzhen, 518172, China and with Shenzhen Research Institute of Big Data, China (e-mail: luozq@cuhk.edu.cn).}
}

%


\maketitle

\begin{abstract}
It remains an open problem to find the optimal configuration of phase shifts under the discrete constraint for intelligent reflecting surface (IRS) in polynomial time. The above problem is widely believed to be difficult because it is not linked to any known combinatorial problems that can be solved efficiently. The branch-and-bound algorithms and the approximation algorithms constitute the best results in this area. Nevertheless, this work shows that the global optimum can actually be reached in linear time on average in terms of the number of reflective elements (REs) of IRS. The main idea is to geometrically interpret the discrete beamforming problem as choosing the optimal point on the unit circle. Although the number of possible combinations of phase shifts grows exponentially with the number of REs, it turns out that there are only a linear number of circular arcs that possibly contain the optimal point. Furthermore, the proposed algorithm can be viewed as a novel approach to a special case of the discrete quadratic program (QP). 
\end{abstract}


\section{Introduction}
\label{sec:Introductio}
\IEEEPARstart{C}{onfiguring} intelligent reflecting surface (IRS) concerns with the choice of a set of ``best'' phase shifts across reflective elements (REs), namely passive beamforming, to boost the signal reception at the target receiver. The existing studies in this area \cite{zhangrui2022survey,OnReview} mainly divide into two categories: those assuming continuous phase shifts to ease optimization or analysis, and those restricting the phase shifts to a given discrete set from a practical standpoint. This work belongs to the latter. We establish a result that seems at first surprising: the discrete IRS beamforming problem has an $O(N)$ algorithm for its solution, where $N$ is the number of REs.

The proposed algorithm stems from a geometric interpretation of the IRS beamforming problem which was first proposed in \cite{yaowen2022jstsp}. While \cite{yaowen2022jstsp} focuses on the binary beamforming case with each phase shift being either $0$ or $\pi$, this work goes further to account for a general $K$-ary case where each phase shift takes on $K$ discrete values. We propose visualizing each possible solution as a point on the unit circle; the point moving around the unit circle captures the different beamforming decisions. The essential power of the above geometric method arises from the following observation: although the number of possible combinations of phase shifts grows exponentially with the number of REs, there are at most a linear number of unit circular arcs that possibly contain the point of the optimal beamforming. Furthermore, the utilities of all these arcs can be computed in linear time. The main idea behind the proposed algorithm is to try out all possible arcs and pick the best, and then recover the optimal phase shifts.

The existing works seldom consider global optimum for the discrete IRS beamforming. Although the IRS technology has been extensively studied over the past few years for many complicated system models, the simplest point-to-point model is still far less well understood. A common practice in the literature is to first solve the relaxed continuous beamforming problem via standard optimization, and then round the solution to the discrete set, e.g., \cite{choi2021oct,changsheng2020july} based on semidefinite relaxation (SDR) and \cite{weiheng2021aug,yongqing2022apr} based on minorization-maximization (MM). For the relaxed continuous beamforming problem, \cite{Gao_DBF_21} enforces the discrete constraint by penalization, but the resulting nonconvex problem is still difficult to deal with. Another popular idea \cite{Yuen_DBF_20,wu_zhang_TCOM20}
is to decide the phase shift for one RE at a time, but the performance of such greedy search is not provable. Some other works \cite{di_song_jsac20,Schober_DBF_21} resort to the branch-and-bound algorithm to obtain the global optimum of the discrete IRS beamforming. But these methods incur exponential time complexities and provide limited insights. For the practical approach, the best result so far in the literature is
to attain an approximation ratio of the global optimum, e.g., \cite{wu_zhang_TCOM20} gives an approximation ratio of $\cos^2(\pi/K)$ in terms of the optimal signal-to-noise ratio (SNR), where $K$ is the number of different values each phase shift can take on, and further \cite{yaowen2022jstsp} improves it to $0.5+0.5\cos(\pi/K)$. Moreover, \cite{Balakrishnan_NSDI20,shuyi2022twc} consider the discrete IRS beamforming in the absence of channel information. To the best of our knowledge, the present work gives the first polynomial time algorithm for the discrete IRS beamforming with provable global optimality.

Furthermore, it is worth mentioning the connection to the quadratic program (QP) like $\max\;\bm x^\top\bm Q \bm x$. The authors of \cite{linglong2021jun} already observed that the discrete IRS beamforming problem at the link level boils down to solving a discrete QP---which is NP-hard in general \cite{hammer2022oct} but can be efficiently solved under certain conditions \cite{tiantian2016jan}, e.g., when each entry of $\bm x$ is selected from a binary set $\{0,1\}$ and $\mathrm{rank}(\bm Q)\le2$. Actually, the quadratic time algorithm proposed in \cite{linglong2021jun} for the binary IRS beamforming is a direct application of the QP method in \cite{tiantian2016jan}. The binary QP without the rank constraint can be solved in polynomial time as well \cite{Allemand2001}. Nevertheless, a generic $K$-ary discrete QP remains an open problem. The proposed algorithm can be recognized as a linear time approach to a special case of the $K$-ary discrete QP.

\section{System Model}
\label{sec:model}

Consider wireless transmission in aid of an IRS. Assume that the IRS comprises $N$ REs, and that the transmitter and receiver have one antenna each. We use $h_0\in\mathbb C$ to denote the direct channel from the transmitter to the receiver, $h^{\text{I}}_n\in\mathbb C$ the channel from the transmitter to the $n$th RE, and $h^{\text{II}}_n\in\mathbb C$ the channel from the $n$th RE to the receiver, so the cascaded reflected channel $h_n\in\mathbb C$ associated with the $n$th RE is
\begin{equation}
\label{hn}
	h_n= h^\text{II}_n h^\text{I}_n,\;\;\text{for}\;n=1,\ldots,N.
\end{equation} 
In the remainder of the paper, the direct and reflected channels are frequently written in a polar form, i.e.,
\begin{equation}
\label{polar}
h_n = \beta_ne^{j\alpha_n},\;\;\text{for}\;n=0,1,\ldots,N,
\end{equation}
with the magnitude $\beta_n>0$ and the phase $\alpha_n\in[0,2\pi)$. In particular, $\beta_n>1$ occurs if IRS can amplify reflection. Assume that the channel phases $\alpha_n$ are uniformly distributed. Moreover, each RE $n$ induces a phase shift $\theta_n$ into its reflected channel $h_n$, which takes on values in a prescribed discrete set
\begin{equation}
\label{Phi}
\Phi_K = \left\{\omega,2\omega,\ldots,K\omega\right\}\;\;\text{where}\; \omega = \frac{2\pi}{K}.
\end{equation}
Thus, for the transmit signal $X\in\mathbb C$, the received signal $Y\in\mathbb C$ is given by
\begin{align}
    Y &= \Bigg(h_0+\sum^N_{n=1}h_ne^{j\theta_n}\Bigg)X+Z,
\end{align}
where $Z\sim\mathcal{CN}(0,\sigma^2)$ is an i.i.d. complex Gaussian noise. 

Denoting the transmit power level by $P$, the SNR can be computed as
\begin{subequations}
\begin{align}
\label{snr}
	\mathsf{SNR} &=\frac{\mathbb E[|Y-Z|^2]}{\mathbb E[|Z|^2]}\\
	&=\frac{P}{\sigma^2}\left|\beta_0e^{j\alpha_0}+\sum^N_{n=1}\beta_ne^{j(\alpha_n+\theta_n)}\right|^2.
\end{align}
\end{subequations}
Similarly, the baseline SNR without IRS can be obtained as
\begin{equation}
	\mathsf{SNR}_0 = \frac{P\beta^2_0}{\sigma^2}.
\end{equation}
To quantify the performance gain due to IRS, we introduce the notion of \emph{SNR boost} as
\begin{subequations}
\begin{align}
\label{SNR_boost}
	f(\bm\theta) &= \frac{\mathsf{SNR}}{\mathsf{SNR}_0}\\
	&=\frac{1}{\beta^2_0}\left|\beta_0e^{j\alpha_0}+\sum^N_{n=1}\beta_ne^{j(\alpha_n+\theta_n)}\right|^2.
\end{align}
\end{subequations}
We now consider the problem of finding the optimal IRS beamforming vector $\bm\theta:=(\theta_1,\ldots,\theta_N)$ that yields the highest SNR boost:
\begin{subequations}
\label{prob:snr}
\begin{align}
\underset{\bm\theta}{\text{maximize}} &\quad f(\bm\theta)
    \label{prob:snr:obj}\\
\text{subject to}&\quad \theta_n\in\Phi_K,\;\;\text{for}\; n=1,\ldots,N.
    \label{prob:snr:cons}
\end{align}    
\end{subequations}
We assume that all the channel information is available. In the above problem, the optimal value of a particular $\theta_n$ heavily depends on how the other $\theta_{m}$'s ($m\ne n$) are chosen. It seems intractable to directly coordinate all these $\theta_n$'s because the size of the solution space $\Phi_K\times\Phi_K\times\ldots\times\Phi_K$ is exponential in $N$. Nevertheless, the rest of the paper shows that the problem \eqref{prob:snr} can be optimally solved in linear time in expectation.

\section{Proposed Method}

\subsection{Geometric Interpretation of Discrete Beamforming}

\begin{figure}[t]
\centering
\centerline{\includegraphics[width=5cm]{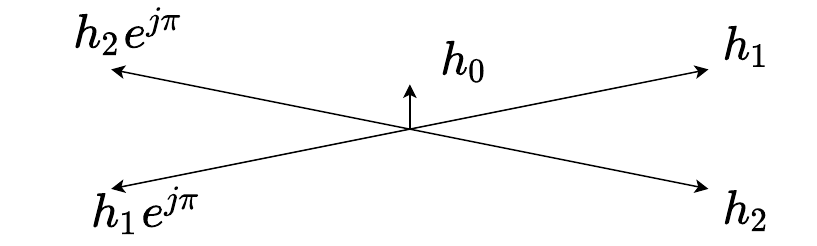}}
\caption{Clearly, it is optimal to set either $\theta_1=\theta_2=0$ or $\theta_1=\theta_2=\pi$ in this example. But we would end up with $\theta_1=0$ and $\theta_2=\pi$ if each $h_n$ is rotated to the closest position to $h_0$, which leads to fairly poor performance.}
\label{fig:bad_case}
\vspace{-1em}
\end{figure}

The problem in \eqref{prob:snr} can be visualized with a complex plane graph where every channel $h_n$ corresponds to a 2-dimensional real vector. Applying the phase shift $\theta_n$ to $h_n$ can be viewed as rotating the vector $h_n$ counterclockwise by an angle of $\theta_n$; recall that the rotation must be discrete because of the constraint $\theta_n\in\Phi_K$. We aim to coordinate the rotations of $(h_1,\ldots,h_N)$ so that the magnitude of their vector addition is maximized. Thus, if $(\theta_1^\star,\ldots,\theta_N^\star)$ is the optimal solution to \eqref{prob:snr}, then the vector addition
\begin{equation}
\label{g}
    g = h_0 + \sum^N_{n=1}h_ne^{j\theta_n^\star}
\end{equation}
has the maximum length. We write the normalized $g$ as
\begin{equation}
\label{mu}
    \mu = \frac{g}{|g|}.
\end{equation}
Notice that the vector $h_0$ is fixed. A naive idea for maximizing $|g|$ is to rotate every $h_n$ to the closest possible position to $h_0$ in the complex plane, namely the \emph{closest point projection (CPP)}. However, CPP is suboptimal and can yield quite bad results, as illustrated in Fig. \ref{fig:bad_case}. The reason is that $h_0$ may considerably deviate from $\mu$, in which case aligning $h_n$ with $h_0$ cannot contribute much to the magnitude of the overall vector addition. Rather, it is always optimal to rotate $h_n$ to the closest possible position to $\mu$, as stated in the following lemma.

\begin{lemma}
\label{lemma:opt_theta}
For an optimal solution $(\theta^\star_1,\ldots,\theta_N^\star)$ to problem \eqref{prob:snr}, each $\theta_n^\star$ must satisfy
 \begin{equation}
 \label{opt_theta}
 \theta^\star_n=\arg\min_{\theta_n\in\Phi_K}\left|\mathrm{Arg}\left(\frac{h_ne^{j\theta_n}}{g}\right)\right|,
 \end{equation}
\end{lemma}
where $\mathrm{Arg}(\cdot)$ is the principal argument of a complex number.
\begin{IEEEproof}
    Suppose that there exists some $\theta'_{n}\in\Phi_K\backslash\{\theta^\star_n\}$ with $\big|\mathrm{Arg}(h_ne^{j\theta'_n}/g)\big|<\big|\mathrm{Arg}(h_ne^{j\theta^\star_n}/g)\big|$. We can then render $h_n$ closer to the addition of the rest vectors by applying a phase shift of $\theta'_n$ to it instead, thereby further increasing $|g|$, which is a contradiction.
\end{IEEEproof}

\begin{figure}[t]
\centering
\centerline{\includegraphics[width=5.5cm]{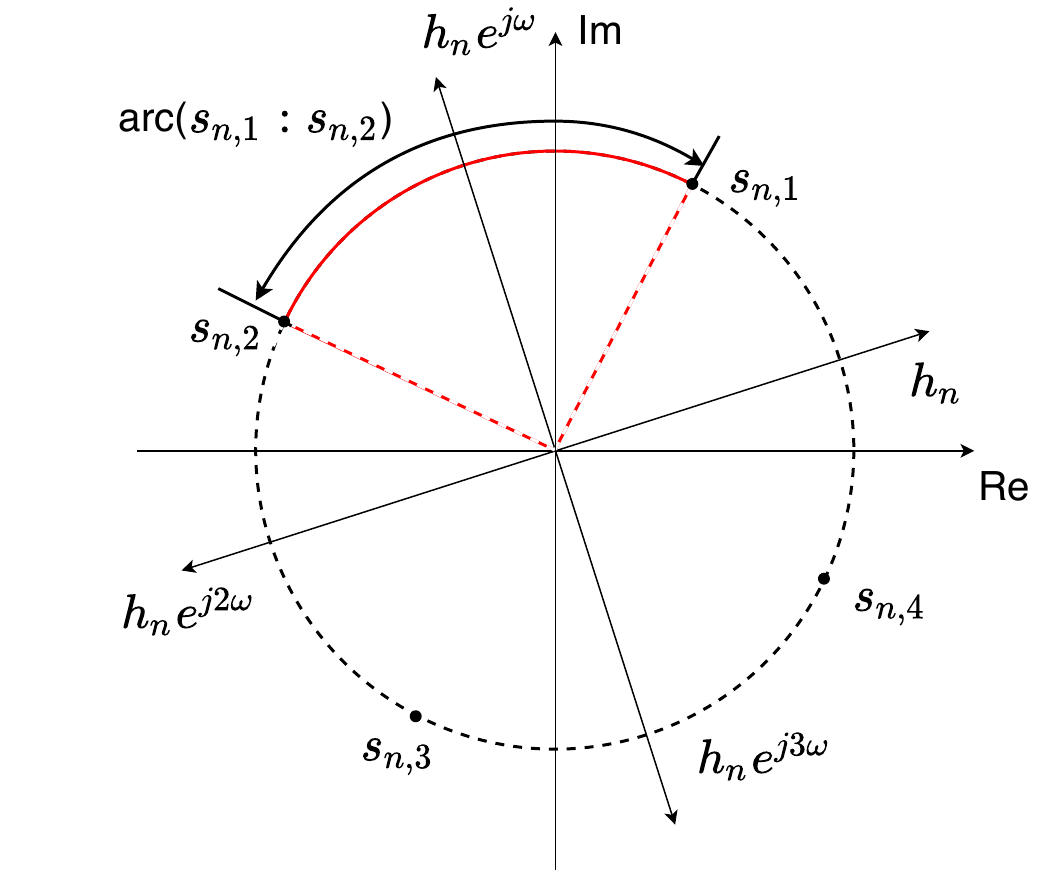}}
\caption{A total of $K$ points $\{s_{nk},k=1,\ldots,K\}$ partition the unit circle into $K$ arcs; $\mathrm{arc}(s_{ni}:s_{n,i+1})$ starts from $s_{ni}$ and ends at $s_{n,i+1}$ in the counterclockwise direction.}
\label{fig:sh}
\vspace{-1em}
\end{figure}

In light of the above lemma, we proceed to decide a range of $\mu$ for which $\theta^\star_n=k\omega$ must hold. First, define a sequence of complex numbers with respect to each $n=1,\ldots,N$ as
\begin{equation}
\label{s}
    s_{nk} = e^{j(\alpha_n+(k-0.5)\omega)},\;\;\text{for}\; k=1,\ldots,K.
\end{equation}
Notice that these complex numbers, when visualized as points in the complex plane, all lie on the unit circle $C$. For any two points $a,b\in C$, we use $\mathrm{arc}(a: b)$ to denote a unit circular arc with $a$ as the initial end and $b$ as the terminal end in the counterclockwise direction; in particular, let $\mathrm{arc}(a: b)$ be an open arc with the two endpoints $a$ and $b$ excluded, as shown in Fig. \ref{fig:sh}. The following proposition follows from Lemma \ref{lemma:opt_theta}.
\begin{proposition}
\label{prop:opt_theta}
A sufficient condition for
$\theta^\star_n=k\omega$ is
\begin{equation}
\label{opt_mu}
    \mu \in\mathrm{arc}(s_{nk}: s_{n,k+1}).
\end{equation}
\end{proposition}
Intuitively, letting $\theta_n=k\omega$ is guaranteed to minimize the gap $\big|(\theta_n+\alpha_n-\angle \mu)\bmod{2\pi}\big|$ whenever $\mu$ lies in its associated arc, and thus $k\omega$ must be optimal according to Lemma \ref{lemma:opt_theta}.

We put together all the complex numbers $\{s_{nk},\,\forall (n,k)\}$, then remove the replicas among them, and further sort out the remaining distrinct numbers according to their phases, i.e.,
\begin{equation}
\label{lambda}
0\le \angle\lambda_1<\angle\lambda_2<\ldots<\angle\lambda_L<2\pi,
\end{equation}
where $\{\lambda_1,\ldots,\lambda_L\}$, $L\le NK$, are the distinct numbers of $\{s_{nk},\,\forall (n,k)\}$. Moreover, we use the following function to retrieve the associated RE indices $n$ from $\lambda_\ell$:
\begin{equation}
    \mathcal N(\lambda_\ell) = \left\{n \,\big|\, s_{nk}=\lambda_\ell\right\}.
\end{equation}
Notice that the distinct points of $(\lambda_1,\ldots,\lambda_L)$ all lie on the unit circle $C$ and partition $C$ into $L$ circular arcs, which are $\mathrm{arc}(\lambda_1:\lambda_2),\ldots,\mathrm{arc}(\lambda_{L-1}:\lambda_{L})$, and $\mathrm{arc}(\lambda_{L}:\lambda_1)$.

The key observation is that if $\mu$ moves around $C$ but within the same $\mathrm{arc}(\lambda_\ell:\lambda_{\ell+1})$, i.e., when $\mu$ stays in the same $\mathrm{arc}(s_{nk}: s_{n,k+1})$ for every $n$, then the optimal solution of $\theta^\star_n$ does not change according to Proposition \ref{prop:opt_theta}. As a result, we conclude that the optimal solution $(\theta^\star_1,\ldots,\theta^\star_N)$ solely depends on which $\mathrm{arc}(\lambda_\ell:\lambda_{\ell+1})$ contains $\mu$. Because $\mu$ is unknown \emph{a priori}, we need to try out all possible arcs and choose the best. The essential power of this exhaustive search approach arises from the fact that there are at most $KN$ arcs, so the complexity is just linear in $N$.

\begin{remark}
\label{remark:2N}
Actually, we can further reduce the number of the considered arcs from $KN$ to $2N$ by using  \cite[Proposition~1]{yaowen2022jstsp}: the optimal overall channel superposition deviates from $h_0$ by less than $2\pi/K$, i.e., $|(\angle\mu-\alpha_0)\bmod 2\pi|<2\pi/K$. Thus, it suffices to consider the two closest arcs to $h_0$ for each $h_n$.
\end{remark}

So far we sketch the main idea behind the proposed method. But there are still two technical challenges to tackle. First, how to break the tie when two phase shifts can both minimize the gap in \eqref{opt_theta}, i.e., when $\mu$ lies right between two arcs rather than in the interior of an arc. Second, how to actually achieve the linear complexity in $N$. Here is a subtle pitfall: if the SNR boost is evaluated for each arc separately, then it requires $O(N^2)$ time to figure the optimal arc. We give the complete description of the proposed method in the sequel.

\subsection{Optimal Solving in Linear Time}

We start by showing that the aforementioned tie-breaking issue will never occur.
\begin{proposition}
\label{prop:endpoint}
    We always have $\mu\ne\lambda_\ell$ for any endpoint $\lambda_\ell$, so $\mu$ must lie in the interior of some $\mathrm{arc}(\lambda_\ell:\lambda_{\ell+1})$.
\end{proposition}
\begin{IEEEproof}
    Suppose that $\mu=\lambda_\ell$ for some $\ell$, so there exists at least one $s_{nk}$ such that $s_{nk}=\mu$. As a result, $\mu$ lies on the boundary between $\mathrm{arc}(s_{n,k-1}:s_{nk})$ and $\mathrm{arc}(s_{nk}:s_{n,k+1})$, and hence we need to break a tie: $\theta_n^\star=(k-1)\omega$ or $\theta_n^\star=k\omega$. Assume that $\theta^\star_n=k\omega$ and let $g'=g-h_ne^{jk\omega}$. It can be easily seen that $g$ must lie between $h_ne^{jk\omega}$ and $g'$. Thus, $g'$ and $h_ne^{j(k-1)\omega}$ must be on the same side of $g$ while $h_ne^{jk\omega}$ lies on the opposite side. Furthermore, because $h_ne^{j(k-1)\omega}$ and $h_ne^{jk\omega}$ are symmetric about $g$, $h_ne^{j(k-1)\omega}$ must be closer to $g'$, and consequently letting $\theta_n=(k-1)\omega$ would increase $|g|$, which forms a contradiction. By symmetry, we can draw a similar contradiction if assuming $\theta^\star_n=(k-1)\omega$ at first. The above proposition is then verified.
\end{IEEEproof}

\begin{remark}
Equipped with Proposition \ref{prop:endpoint}, we can strengthen Proposition \ref{prop:opt_theta} to state that \eqref{opt_mu} is a necessary and sufficient condition for $\theta^\star_n=k\omega$.
\end{remark}

Because of Proposition \ref{prop:endpoint}, we can restrict the possible positions of $\mu$ to the interior of arcs. We seek the optimal arc that maximizes $|g|$. Let us start from $\mathrm{arc}(\lambda_1:\lambda_2)$ and move $\mu$ around the unit circle in the counterclockwise direction. As $\mu$ enters the $(\ell+1)$th arc from the previous $\ell$th arc, i.e.,
\begin{equation}
    \mathrm{arc}(\lambda_\ell:\lambda_{\ell+1}) \rightarrow \mathrm{arc}(\lambda_{\ell+1}:\lambda_{\ell+2}),
\end{equation}
all those $\theta_n$ with $n\in\mathcal N(\lambda_{\ell+1})$ are changed accordingly because they no longer minimize the phase gaps $\left|\mathrm{Arg}\left(\frac{h_ne^{j\theta_n}}{g}\right)\right|$ in \eqref{opt_theta}. These $\theta_n$'s are optimally updated as
\begin{equation}
\label{update_theta}
\theta_n \rightarrow \theta_n+\omega\;\;\text{for all}\; n\in\mathcal N(\lambda_{\ell+1}).
\end{equation}
Importantly, notice that the overall vector addition $g$ is effected only by those newly updated $\theta_n$ with $n\in\mathcal{N}(\lambda_{\ell+1})$ in \eqref{update_theta}. Thus, the overall channel superposition in the new arc, denoted by $g_{\ell+1}$, can be efficiently obtained from the previous $g_{\ell}$ as
\begin{equation}
\label{update_g}
    g_{\ell+1} = g_{\ell} + \sum_{n\in\mathcal N(\lambda_{\ell+1})}\left(h_ne^{j\theta_n}-h_ne^{j(\theta_n-\omega)}\right).
\end{equation}
The above recursive computation of $g_{\ell+1}$ plays a key role in our linear time algorithm, as shown in the proof of Theorem \ref{theorem:opt} in what follows. Otherwise, if each $g_{\ell}$ is evaluated according to \eqref{g} separately, then the complexity would raise to $O(N^2)$. Algorithm \ref{alg} summarizes the steps. 

\begin{algorithm}[t]
\caption{Linear-Time Optimal Beamforming for IRS}
\label{alg}
\begin{algorithmic}[1]
\State\textbf{Initialization:} Compute $\{s_{nk}\}$ and sort $\{\lambda_1,\ldots,\lambda_L\}$ according to \eqref{s} and \eqref{lambda}.\\
Find $L$ arcs that possibly contain $\mu$ as in Remark \ref{remark:2N}.\\
Compute each $\theta_n$ according to \eqref{opt_theta}; obtain $g_1$ in \eqref{g}.
\For{each possible $\mathrm{arc}(\lambda_\ell:\lambda_{\ell+1})$}
    \State Update $\theta_{n}\rightarrow\theta_n+\omega$ for each $n\in\mathcal N(\lambda_\ell)$.
    \State Compute $g_\ell$ based on $g_{\ell-1}$ according to \eqref{update_g}.
\EndFor
\State Find $\ell_\text{OPT}=\arg\max_\ell \{g_\ell\}$.\\
Place $\mu$ anywhere in the interior of $\mathrm{arc}(\lambda_{\ell_\text{OPT}}:\lambda_{\ell_\text{OPT}+1})$.\\
Compute each optimal $\theta^\star_n$ according to \eqref{opt_theta}.
\end{algorithmic}
\end{algorithm}    

\begin{theorem}
\label{theorem:opt}
 Algorithm \ref{alg} yields the global optimal solution $(\theta^\star_1,\ldots,\theta^\star_N)$ to problem \eqref{prob:snr} in $O(N)$ time in expectation.
\end{theorem}
\begin{IEEEproof}
    The global optimality of Algorithm \ref{alg} is evident because each $\theta_n$ is optimally decided as in Proposition \ref{prop:opt_theta} and all the possible arcs have been considered for $\mu$.
    
    We now analyze the complexity of Algorithm \ref{alg}. Recall that we have assumed that the channel phases $\alpha_n$'s are uniformly distributed on $[0,2\pi)$, and thus $\lambda_\ell$'s are uniformly distributed on the unit circle. By the binning method \cite{TheAlgorithm}, it takes $O(N)$ time on average to sort the $\lambda_\ell$'s in step 1 of Algorithm \ref{alg}; we remark that the worst-case complexity is $O(N^2)$. Regarding the for-loop from step 5 to step 8, for the iteration of $\ell$, \eqref{update_theta} and \eqref{update_g} incur $O(|\mathcal{N}(\lambda_\ell)|)$ each, so the entire for-loop incurs $\sum^L_{\ell=2}O(|\mathcal{N}(\lambda_\ell)|)=O(N)$. Moreover, we see immediately that the complexity of the rest steps is linear in $N$. Thus, the overall complexity is $O(N)$ on average.
\end{IEEEproof}

\begin{figure}[t]
\centering
\centerline{\includegraphics[width=8cm]{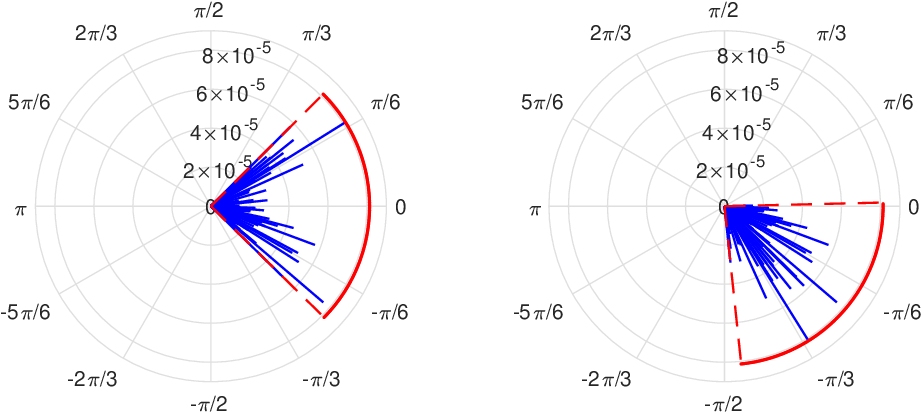}}
\caption{Plot of the rotated channels $h_ne^{j\theta_n}$ when $N=100$ and $K=4$. Algorithm \ref{alg} (right) outperforms CPP (left) by making channels more clustered.}
\label{fig:polar}
\vspace{-1em}
\end{figure}

The average complexity of Algorithm \ref{alg} would raise to $O(NK)$ if we try out all the arcs on the unit circle. Moreover, if the channel phases are not uniformly distributed on $[0,2\pi)$, then we may sort out $\lambda_\ell$'s in $N\log(N)$ time on average by using the quick sort instead in step 1 of Algorithm \ref{alg}. Fig.~\ref{fig:polar} illustrates the gain of Algorithm \ref{alg} over CPP.

\begin{remark}
    Algorithm \ref{alg} can be readily applied to a complex $K$-ary QP: $\mathrm{maximize}_{\bm x}\;\bm x^\top\bm Q\bm x$ subject to $x_i\in\{1,\omega,\ldots,\omega^{K-1}\}$ whenever $\mathrm{rank}(\bm Q)=1$ over $\mathbb C$, i.e., when $\bm Q$ can be decomposed as $\bm Q=\bm v^H\bm v$ for some complex vector $\bm v$.
\end{remark}




\section{Simulation Results}

We now compare the proposed algorithm with the existing methods for the discrete IRS beamforming in some numerical examples. The transmit power is 30 dBm and the background noise power is $-90$ dBm. The channel model follows \cite{wu_zhang_TCOM20,Yuen_DBF_20,wy_AI_21} as specified in the following. The direct channel is modeled as 
$h_0=10^{-\textsf{PL}_0/20}\times\zeta_0$ where $\textsf{PL}_0=32.6+36.7\log_{10}(d_0)$ is the pathloss (in dB) between the transmitter and the receiver which are $d_0$ meters apart, and $\zeta_0$ is the Rayleigh fading component drawn i.i.d. from the Gaussian distribution $\mathcal{CN}(0,1)$. The cascaded reflected channel is modeled as
$h_n=10^{-(\textsf{PL}_1+\textsf{PL}_2)/20}\times\zeta_{n1}\zeta_{n2}$
where $\textsf{PL}_1$ and $\textsf{PL}_2$ are computed as $\textsf{PL}_i=30+22\log_{10}(d)$, $i\in\{1,2\}$, with $d$ in meters denoting the transmitter-to-IRS distance and the IRS-to-receiver distance, respectively. The Rayleigh fading components $\zeta_{n1}$ and $\zeta_{n2}$ are drawn from the Gaussian distribution $\mathcal{CN}(0,1)$ independently across $n=1,\ldots,N$. The locations of the transmitter, IRS, and receiver are denoted by the 3-dimensional coordinate vectors $(50,-200,20)$, $(-2,-1,0)$, and $(0,0,0)$ in meters, respectively. We compare Algorithm \ref{alg} with the exhaustive search when $N=10$, both of which achieve the global optimum. For a larger number of REs, say $N=100$, we consider the following two baseline methods for comparison purpose:
\begin{itemize}
    \item \emph{Closest Point Projection (CPP) \cite{wu_zhang_TCOM20}:} Round the continuous solution $\theta_n=\alpha_0-\alpha_n$ to the closest point in $\Phi_K$.
    \item \emph{Block Coordinate Descent (BCD) \cite{Yuen_DBF_20}:} Optimize the phase shift for one RE at a time.
\end{itemize}

\begin{figure}[t]
    \centering
    \includegraphics[width=9cm]{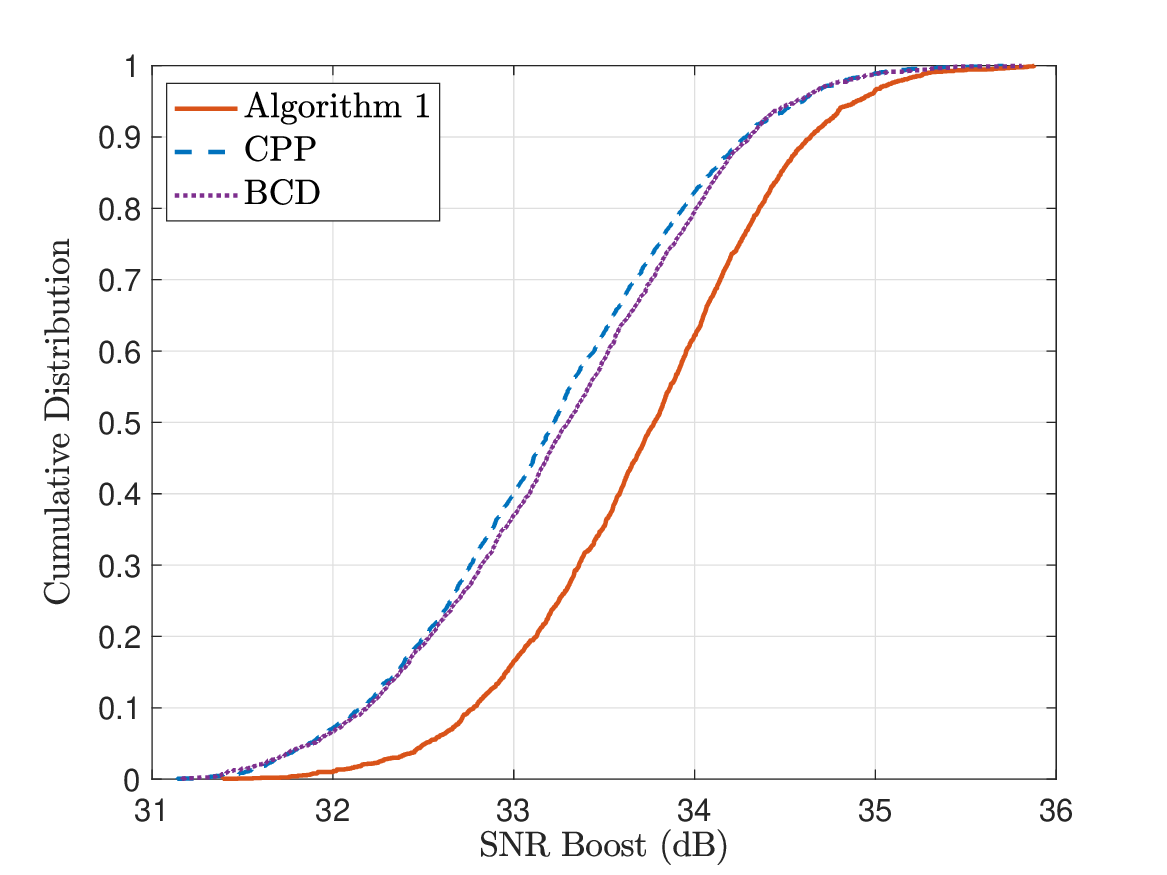}
    \caption{Cumulative distribution of SNR boost when $K=2$.}
    \label{N100K2U1}
    \vspace{-1em}
\end{figure}

\begin{figure}[t]
    \centering
    \includegraphics[width=9cm]{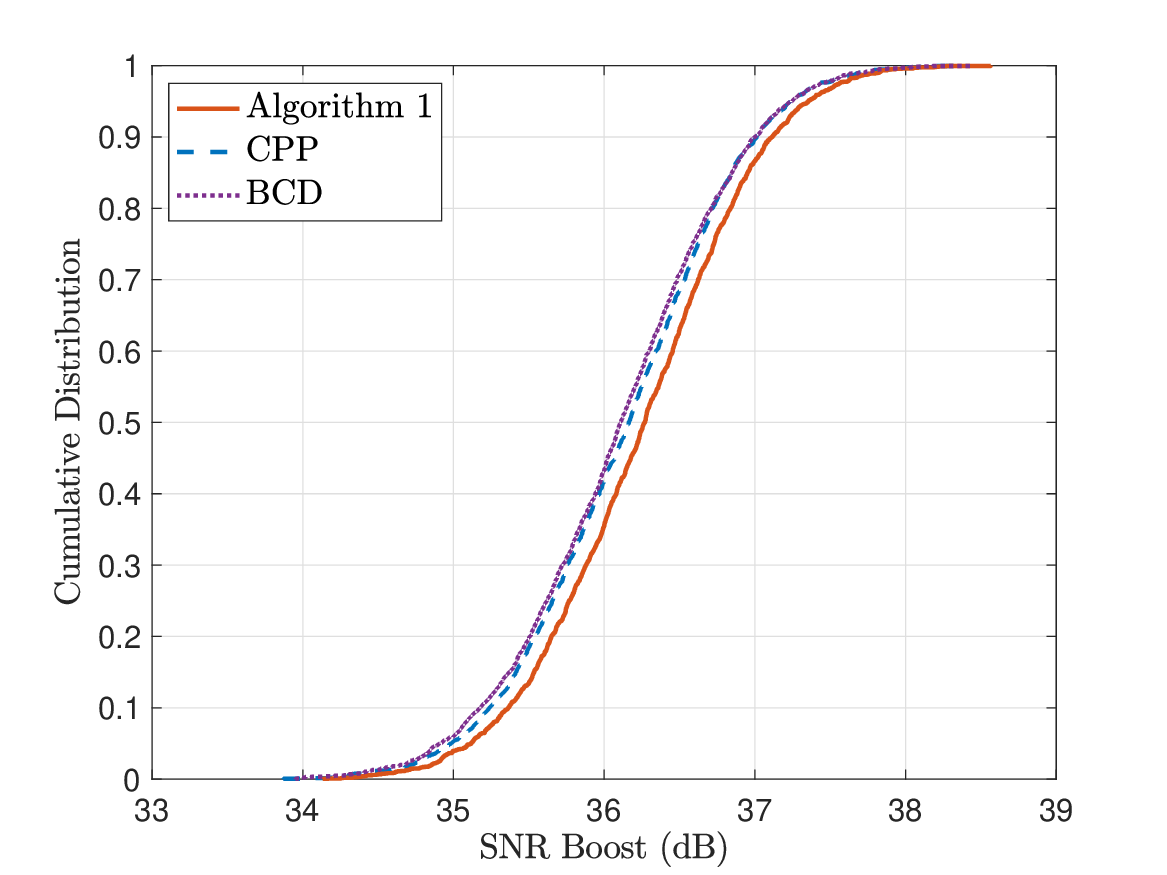}
    \caption{Cumulative distribution of SNR boost when $K=4$.}
    \label{N100K3U1}
    \vspace{-1em}
\end{figure}

Fig.~\ref{N100K2U1} shows the cumulative distribution of the SNR boost when the number of phase shift choices $K=2$. While the proposed algorithm is guaranteed to attain the global optimum, CPP and BCD are both suboptimal. Fig.~\ref{N100K3U1} further shows the case of $K=4$. CPP and BCD are still suboptimal but the gaps become smaller. Actually, the performance of CPP and BCD are sensitive to the choice of channel model. For instance, as previously shown in Fig.~\ref{fig:bad_case}, the gap between CPP and the global optimum can be arbitrarily large in certain scenarios. In particular, if the provable ultrareliable transmission is of crucial importance, then the proposed algorithm is a much better choice than the existing methods. 

\begin{figure}[t]
    \centering
    \includegraphics[width=9cm]{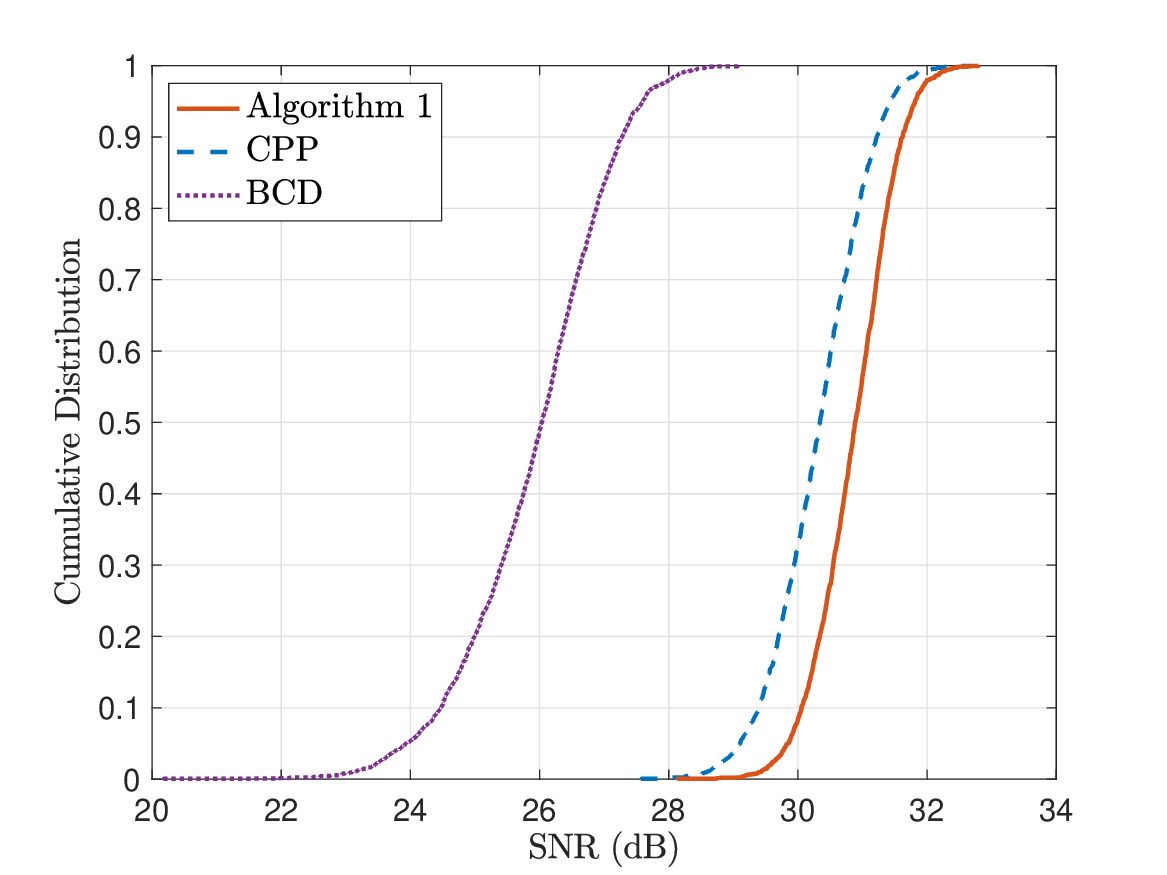}
    \caption{Cumulative distribution of min SNR across 4 users when $K=2$.}
    \label{fig:multiuser}
    \vspace{-1em}
\end{figure}

We further consider multiple receivers when $K=2$. In addition to the existing receiver at $(0,0,0)$, three more receivers are located at $(0,1,0)$, $(1,0,0)$, and $(1,1,0)$, respectively. Assume that a common message is intended to the four receivers, namely the broadcast channel. We propose running Algorithm \ref{alg} (or CPP, or BCD) for each receiver and then choose the best $\{\theta_n\}$ to maximize the lowest SNR across the four receivers. Thus, the time complexity of this extended Algorithm \ref{alg} is $O(NU)$ where $U$ refers to the number of receivers. The resulting cumulative distribution of the lowest SNR by the different methods is displayed in Fig.~\ref{fig:multiuser}. Again, the proposed method outperforms all the other methods. Note that BCD has quite bad performance in this multi-user case.

\section{Conclusion}

We propose a linear time algorithm that is guaranteed to yield the globally optimal configuration of discrete phase shifts for IRS in order to maximize the SNR boost, whereas the best results in the literature are branch-and-bound algorithms and the approximation algorithms. The proposed method provides insights into how the phase shifts should be coordinated across the REs from a geometric perspective. Moreover, this work sheds light on how the notoriously difficult problem of discrete QP can be solved for a special case.

\bibliographystyle{IEEEtran}     
\bibliography{IEEEabrv,strings}     

\begin{thebibliography}{10}
\providecommand{\url}[1]{#1}
\csname url@samestyle\endcsname
\providecommand{\newblock}{\relax}
\providecommand{\bibinfo}[2]{#2}
\providecommand{\BIBentrySTDinterwordspacing}{\spaceskip=0pt\relax}
\providecommand{\BIBentryALTinterwordstretchfactor}{4}
\providecommand{\BIBentryALTinterwordspacing}{\spaceskip=\fontdimen2\font plus
\BIBentryALTinterwordstretchfactor\fontdimen3\font minus
  \fontdimen4\font\relax}
\providecommand{\BIBforeignlanguage}[2]{{%
\expandafter\ifx\csname l@#1\endcsname\relax
\typeout{** WARNING: IEEEtran.bst: No hyphenation pattern has been}%
\typeout{** loaded for the language `#1'. Using the pattern for}%
\typeout{** the default language instead.}%
\else
\language=\csname l@#1\endcsname
\fi
#2}}
\providecommand{\BIBdecl}{\relax}
\BIBdecl

\bibitem{zhangrui2022survey}
B.~Zheng, C.~You, W.~Mei, and R.~Zhang, ``A survey on channel estimation and
  practical passive beamforming design for intelligent reflecting surface aided
  wireless communications,'' \emph{{IEEE} Commun. Surveys Tuts.}, vol.~24,
  no.~2, pp. 1035--1071, Feb. 2022.

\bibitem{OnReview}
C.~Pan, G.~Zhou, K.~Zhi, S.~Hong, T.~Wu, Y.~Pan, H.~Ren, M.~D. Renzo, S.~A.
  Lee, R.~Zhang, and A.~Y. Zhang, ``An overview of signal processing techniques
  for {RIS/IRS}-aided wireless systems,'' \emph{{IEEE} J. Sel. Topics Signal
  Process.}, vol.~16, no.~5, pp. 883--917, Aug. 2022.

\bibitem{yaowen2022jstsp}
Y.~Zhang, K.~Shen, S.~Ren, X.~Li, X.~Chen, and Z.-Q. Luo, ``Configuring
  intelligent reflecting surface with performance guarantees: {O}ptimal
  beamforming,'' \emph{{IEEE} J. Sel. Topics Signal Process.}, vol.~16, no.~5,
  pp. 967--979, 2022.

\bibitem{choi2021oct}
J.~Choi, L.~Cantos, and Y.~Kim, ``Low-complexity passive beamforming for
  {IRS}-aided uplink {NOMA},'' in \emph{Int. Conf. Inf. Commun. Technol.
  Convergence (ICTC)}, Oct. 2021, pp. 831--833.

\bibitem{changsheng2020july}
C.~You, B.~Zheng, and R.~Zhang, ``Channel estimation and passive beamforming
  for intelligent reflecting surface: Discrete phase shift and progressive
  refinement,'' \emph{{IEEE} J. Sel. Areas Commun.}, vol.~38, no.~11, pp.
  2604--2620, Jul. 2020.

\bibitem{weiheng2021aug}
W.~Jiang, B.~Chen, J.~Zhao, Z.~Xiong, and Z.~Ding, ``Joint active and passive
  beamforming design for the {IRS}-assisted {MIMOME-OFDM} secure
  communications,'' \emph{{IEEE} Trans. Veh. Technol.}, vol.~70, no.~10, pp.
  10\,369--10\,381, Aug. 2021.

\bibitem{yongqing2022apr}
Y.~Xu, Y.~Li, and J.~Wu, ``Weighted sum-rate outage probability constrained
  transmission design for {IRS}-enhanced communication,'' in \emph{Proc. {IEEE}
  Wireless Commun. Netw. Conf. (WCNC)}, Apr. 2022, pp. 1081--1086.

\bibitem{Gao_DBF_21}
L.~You, J.~Xiong, D.~W.~K. Ng, C.~Yuen, W.~Wang, and X.~Gao, ``Energy
  efficiency and spectral efficiency tradeoff in {RIS}-aided multiuser {MIMO}
  uplink transmission,'' \emph{{IEEE} Trans. Signal Process.}, vol.~69, pp.
  1407--1421, Mar. 2021.

\bibitem{Yuen_DBF_20}
S.~Abeywickrama, R.~Zhang, Q.~Wu, and C.~Yuen, ``Intelligent reflecting
  surface: {P}ractical phase shift model and beamforming optimization,''
  \emph{{IEEE} Trans. Commun.}, vol.~68, no.~9, pp. 5849--5863, Sep. 2020.

\bibitem{wu_zhang_TCOM20}
Q.~Wu and R.~Zhang, ``Beamforming optimization for wireless networks aided by
  intelligent reflecting surface with discrete phase shifts,'' \emph{{IEEE}
  Trans. Commun.}, vol.~68, no.~3, pp. 1838--1851, Mar. 2020.

\bibitem{di_song_jsac20}
B.~Di, H.~Zhang, L.~Song, Y.~Li, Z.~Han, and H.~V. Poor, ``Hybrid beamforming
  for reconfigurable intelligent surface based multi-user communications:
  {Achievable} rates with limited discrete phase shifts,'' \emph{{IEEE} J. Sel.
  Areas Commun.}, vol.~38, no.~8, pp. 1809--1822, Aug. 2020.

\bibitem{Schober_DBF_21}
X.~Yu, D.~Xu, and R.~Schober, ``Optimal beamforming for {MISO} communications
  via intelligent reflecting surfaces,'' in \emph{IEEE Int. Workshop Signal
  Process. Advances Wireless Commun. (SPAWC)}, May 2020.

\bibitem{Balakrishnan_NSDI20}
V.~Arun and H.~Balakrishnan, ``{RFocus}: {B}eamforming using thousands of
  passive antennas,'' in \emph{{USENIX} Symp. Netw. Sys. Design Implementation
  ({NSDI})}, Feb. 2020, pp. 1047--1061.

\bibitem{shuyi2022twc}
S.~Ren, K.~Shen, Y.~Zhang, X.~Li, X.~Chen, and Z.-Q. Luo, ``Configuring
  intelligent reflecting surface with performance guarantees: Blind
  beamforming,'' \emph{IEEE Trans. Wireless Commun.}, to be published.

\bibitem{linglong2021jun}
Z.~Zhang and L.~Dai, ``A joint precoding framework for wideband reconfigurable
  intelligent surface-aided cell-free network,'' \emph{{IEEE} Trans. Signal
  Process.}, vol.~69, pp. 4085--4101, Jun. 2021.

\bibitem{hammer2022oct}
P.~L. Hammer, P.~Hansen, P.~M. Pardalos, and D.~J.~R. Jr., ``Maximizing the
  product of two linear functions in 0-1 variables,'' \emph{Optim.}, vol.~51,
  no.~3, pp. 511--537, Oct. 2002.

\bibitem{tiantian2016jan}
T.~Nie, S.~Fang, Z.~Deng, and J.~E. Lavery, ``On linear conic relaxation of
  discrete quadratic programs,'' \emph{Optimi. Methods Softw.}, vol.~31, no.~4,
  pp. 737--754, Jan. 2016.

\bibitem{Allemand2001}
K.~Allemand, K.~Fukuda, T.~M. Liebling, and E.~Steiner, ``A polynomial case of
  unconstrained zero-one quadratic optimization,'' \emph{Mathematical
  Program.}, vol.~91, no.~1, pp. 49--52, Oct. 2001.

\bibitem{TheAlgorithm}
T.~H. Cormen, C.~E. Leiserson, R.~L. Rivest, and C.~Stein, \emph{Introduction
  to Algorithms, 3rd Edition}.\hskip 1em plus 0.5em minus 0.4em\relax {MIT}
  Press, 2009.

\bibitem{wy_AI_21}
T.~Jiang, H.~V. Cheng, and W.~Yu, ``Learning to reflect and to beamform for
  intelligent reflecting surface with implicit channel estimation,''
  \emph{{IEEE} J. Sel. Areas Commun.}, vol.~39, no.~6, pp. 1913--1945, Jul.
  2021.

\end{thebibliography}

\end{document}